\documentstyle[aps,prl]{revtex}
\draft

\def\lsim{\mathrel{\raise.3ex\hbox{$<$\kern-.75em\lower1ex\hbox{$\sim$}}}}
\def\gsim{\mathrel{\raise.3ex\hbox{$>$\kern-.75em\lower1ex\hbox{$\sim$}}}}

\begin{document}

\twocolumn[\hsize\textwidth\columnwidth\hsize\csname
@twocolumnfalse\endcsname

\title {Probing Quantum Decoherence with High-Energy Neutrinos}
\author{Dan Hooper$^1$, Dean Morgan$^2$ and Elizabeth Winstanley$^2$}
\address{
$^1$ Theoretical Astrophysics, University of Oxford, Oxford, UK;\\
$^2$ Department of Applied Mathematics, University of Sheffield,
Sheffield, UK}
\date{\today}

\maketitle

\begin{abstract}

We consider the prospects for observing the effects of quantum
decoherence in high-energy (TeV-PeV) neutrinos from astrophysical
sources. In particular, we study Galactic sources of electron
anti-neutrinos produced in the decay of ultra-high energy
neutrons. We find that next generation neutrino telescopes should
be capable of placing limits on quantum decoherence effects over
multi-kiloparsec baselines, surpassing current bounds by a factor
of $10^{12}$ to $10^{33}$, depending on the model considered.

\end{abstract}

]

\section{Introduction}

Within the context of standard quantum mechanics, a pure state
will never oscillate into a superposition or mixture of states. If
quantum fluctuations of the gravitational field are considered,
however, this may not be the case. Microscopic black holes forming
for short periods of time can lead to a loss of quantum
information, potentially converting a pure state into a mixture or
superposition of quantum states \cite{hawking,QMV}. If evidence of
this effect, called quantum decoherence, were observed, it could
reveal clues about the quantum nature of gravity with incredible
implications for string theory, cosmology and particle physics.

Neutrinos provide a promising sector for observing the
effects of quantum decoherence. Although atmospheric, solar and supernova neutrinos have been previously studied in this context
\cite{lorentz,atmsolar}, high-energy neutrinos produced in distant
astrophysical sources may also be used to search for these
effects. Being weakly interacting, neutrinos can travel very long
distances without scattering. Neutrinos generated in distant
cosmic accelerators provide us with an opportunity to observe
particles which have travelled from elsewhere in our Galaxy
(kilo-parsecs), from nearby galaxies (mega-parsecs) or from
cosmological scales (giga-parsecs). Neutrino measurements over such long baselines
have not yet been conducted and would represent a major step forward in sensitivity to quantum decoherence effects.

As neutrinos propagate, the effects of quantum decoherence would
alter the ratios of their flavors toward the values, $\nu_e:\nu_{\mu}:\nu_{\tau} \cong \frac{1}{3}:\frac{1}{3}:\frac{1}{3}$, regardless of their
initial flavor content. If a flux of neutrinos were to be observed
from a astrophysical source with a ratio of flavors differing from
$\frac{1}{3}:\frac{1}{3}:\frac{1}{3}$, strong constraints could be placed on the scale of quantum
decoherence.

The sources of high-energy cosmic neutrinos most widely studied accelerate protons which interact with photons producing
charged and neutral pions. The charged pions then decay producing
neutrinos in the chain, $\pi^+ \rightarrow \mu^+ \nu_{\mu}
\rightarrow e^+ \bar{\nu_{\mu}} \nu_e \nu_{\mu}$. This initial
flavor ratio ($\frac{1}{3}:\frac{2}{3}:0$) is modified, however, as the neutrinos
undergo oscillations. As a result, cosmic neutrinos produced in
pion decay reach the Earth with flavors very near the ratio of $\frac{1}{3}:\frac{1}{3}:\frac{1}{3}$;
indistinguishable from the signature of quantum decoherence~\cite{equal}.

To test for the effects of quantum decoherence, therefore, a
different type of cosmic neutrino source is needed. Such a source
must produce neutrinos of flavors which do not follow the
standard pion decay ratios. Some possible sources could be neutrinos produced in the annihilations or decays of dark matter
particles \cite{darkmatter}, neutrinos generated as a result of
models of top-down origin of the highest energy cosmic rays
\cite{topdownnus} or neutrinos produced in neutron decays. In
this letter, we study the third of these possibilities.

Neutrons are an interesting source of (anti)neutrinos for our purposes because they
produce neutrinos in only the electron flavor, {\it i.e.} $n \rightarrow
p\, e^- \bar{\nu_e}$. After standard oscillations, this purely electron
anti-neutrino beam converts approximately as 1:0:0 $\rightarrow$
0.56:0.24:0.20. If a cosmic neutrino source were to be found with
such a ratio, this could be used to constrain the scale of quantum
decoherence in the neutrino sector. Alternatively, if we could be
confident that a source produced neutrinos mostly via neutron
decay, the observation of equal quantities of each neutrino flavor
from such a source could potentially constitute a discovery of
quantum decoherence effects.

\section{Cosmic Neutrinos From Neutron Decay}

There are several potentially viable mechanisms for the production
of high-energy cosmic neutrons. For example, ultra-high energy
nuclei accelerated in compact objects can undergo
photo-disintegration through interactions with infrared and
microwave photons, breaking into neutrons and protons
\cite{stecker,halzenneutron}. For example, an accelerated Iron nucleus can
undergo the typical reaction: $^{56}$Fe + $\gamma \rightarrow
^{55}$Fe + $n$. In the region near a cosmic accelerator, sufficiently dense photon fields
may be present to induce such interactions.
Alternatively, neutrons could be produced in charge exchange
interactions of accelerated protons with ambient protons, $p + p
\rightarrow n + X$ \cite{neutrongc}.

There is accumulating evidence for a substantial neutron component
in the cosmic ray spectrum at energies around $10^{18}$ eV. The
Akeno Giant Air Shower Array (AGASA) has reported an anisotropy
correlated to the Galactic Plane at 4 to 4.5$\sigma$ significance.
This excess constitutes about 4$\%$ of the total flux and appears
to be concentrated around the locations of the Galactic Center and
the Cygnus region \cite{agasa}. The reanalysis of data taken by
the Sydney University Giant Airshower Recorder (SUGAR) has also
found evidence for the anisotropy \cite{sugar}. Thirdly, the Fly's
Eye Collaboration has reported an excess along the Galactic Plane
at the 3.2$\sigma$ level \cite{flyseye}. Recently, multi-TeV
gamma-rays have been observed from the vicinity of the Galactic
center \cite{hess}, which may be associated with the cosmic ray
excess from this region \cite{neutrongc,gc}.

If these cosmic ray sources are truly point-like, as the evidence
is beginning to suggest, these events will be somewhat difficult
to reconcile with charged cosmic rays, such as protons, which
would be deflected in the Galactic magnetic fields. With this
motivation, it has been argued that the excess of cosmic rays
around $10^{18}$ eV seen from the Cygnus and Galactic Center
regions (or thereabouts) are neutrons generated in the
photo-disintegration of heavy nuclei or in $pp$ collisions
\cite{halzenneutron,neutrongc}. Energetic neutrons, with a decay
length of $c \gamma_n \tau_n \sim 10 \, \rm{kpc} \times
(E_n/\rm{EeV})$, can reach Earth from such sources only at
energies above about $10^{18}$ eV. The fact that this is the same
energy at which the cosmic ray anisotropies appear is quite
suggestive. Therefore, Galactic sources such as a microquasar,
Cygnus X-3 or the Cygnus-OB2 cluster in the Cygnus region or the
supernova remnant Sgr A East or the supermassive black hole at Sgr A$^*$ in the Galactic Center region (which
are each on the order of 10 kpc from Earth) could be the source of EeV
neutrons. Neutrons would also be generated in these sources at
lower energies which would decay in flight, generating a rich
source of electron anti-neutrinos at PeV energies and below.

Normalizing the neutron flux to the 4$\%$ anisotropic component observed by
AGASA, the authors of Ref.~\cite{halzenneutron} conservatively
estimate an integrated anti-neutrino flux of $\sim 2 \times
10^{-11} \, \rm{cm}^{-2} \rm{s}^{-1}$ above 1 TeV from the Cygnus
region. A similar flux would be expected from the Galactic Center
region. Note that the Waxman-Bahcall bound \cite{wb} on
high-energy neutrino fluxes does not apply here due to their
Galactic nature.

\section{Quantum Decoherence Effects}

We will adopt a simple phenomenological approach to modelling the
effects of quantum decoherence. This is possible, in part, because the effects of quantum decoherence become very simple
when very long propagation distances are considered.
Averaging the sines and cosines which appear in the general
expression, we arrive at the approximate probability of a neutrino
transitioning from state $a$ to state $b$:
\begin{eqnarray}
P[\nu_a \rightarrow \nu_b]&=& \frac{1}{3} + e^{-2 \alpha L} \bigg[\frac{1}{2}(U_{a1}^2 - U_{a2}^2)(U_{b1}^2 - U_{b2}^2) \nonumber \\
&+&\frac{1}{6}(U_{a1}^2 + U_{a2}^2 - 2 U_{a3}^2)(U_{b1}^2 + U_{b2}^2 - 2 U_{b3}^2)\bigg].
\end{eqnarray}
Here, the $U$'s are elements of the standard neutrino mixing
matrix, $\alpha$ is a quantum decoherence parameter, which can
be a constant or a function of energy and $L$ is the distance the
neutrino has travelled. We have used $\Delta m_{21}^2 = 7.2 \times
10^{-5} \, \rm{eV}^2$, $\Delta m_{32}^2 = 2.6 \times 10^{-3} \,
\rm{eV}^2$ and have assumed the normal mass hierarchy. Inserting
the measured neutrino mass splittings and mixing angles, this
further reduces to:
\begin{eqnarray}
\label{prob}
P[\nu_e \rightarrow \nu_e]&=& \frac{1}{3} + 0.228 e^{-2 \alpha L}, \nonumber \\
P[\nu_e \rightarrow \nu_{\mu}]&=& \frac{1}{3} - 0.097 e^{-2 \alpha L}, \nonumber \\
P[\nu_e \rightarrow \nu_{\tau}]&=& \frac{1}{3} - 0.130 e^{-2 \alpha L}, \nonumber \\
P[\nu_{\mu} \rightarrow \nu_{\mu}]&=& \frac{1}{3} + 0.044 e^{-2 \alpha L}, \nonumber \\
P[\nu_{\mu} \rightarrow \nu_{\tau}]&=& \frac{1}{3} + 0.053 e^{-2 \alpha L}, \nonumber \\
P[\nu_{\tau} \rightarrow \nu_{\tau}]&=& \frac{1}{3} + 0.077 e^{-2 \alpha L}.
\end{eqnarray}
In terms of these probabilities, we can write the ratios of neutrino flavors observed
\begin{eqnarray}
R_{\nu_e} &=& (P[\nu_e \rightarrow \nu_e] \Phi_{\nu_e} + P[\nu_{\mu} \rightarrow \nu_e] \Phi_{\nu_{\mu}} \nonumber \\ &+& P[\nu_{\tau} \rightarrow \nu_e] \Phi_{\nu_{\tau}})/\Phi_{\rm{tot}}, \nonumber \\
R_{\nu_{\mu}} &=& (P[\nu_e \rightarrow \nu_{\mu}] \Phi_{\nu_e} + P[\nu_{\mu} \rightarrow \nu_{\mu}] \Phi_{\nu_{\mu}} \nonumber \\ &+& P[\nu_{\tau} \rightarrow \nu_{\mu}] \Phi_{\nu_{\tau}})/\Phi_{\rm{tot}}, \nonumber \\
R_{\nu_{\tau}} &=& (P[\nu_e \rightarrow \nu_{\tau}] \Phi_{\nu_e} +
P[\nu_{\mu} \rightarrow \nu_{\tau}] \Phi_{\nu_{\mu}} \nonumber \\
&+& P[\nu_{\tau} \rightarrow \nu_{\tau}]
\Phi_{\nu_{\tau}})/\Phi_{\rm{tot}},
\label{ratios}
\end{eqnarray}
where the $\Phi$'s denote the respective fluxes emitted at the source. In the case of neutrinos from neutron decay, only $\Phi_{\nu_{e}}$ is non-zero. Notice that if we insert $\Phi_{\nu_{e}}$:$\Phi_{\nu_{\mu}}$:$\Phi_{\nu_{\tau}}=1:0:0$ and $\alpha=0$ (no quantum decoherence) into equations~\ref{ratios} and \ref{prob}, respectively, we find the observed ratios, 0.56:0.24:0.20, described in the introduction.

At this point, we will consider some phenomenological models:
\begin{itemize}


\item{{\bf Energy Independent Model}} \\
If we assume that the quantum decoherence parameter, $\alpha$, is
independent of energy, we find that the probabilities of
Eq.~\ref{prob} all approach their asymptotic value of 1/3 for
distances of $L \gg \alpha^{-1}$. With a source on the order of 10
kiloparsecs distant, values of $\alpha$ on the order of $\sim
10^{-21}$ m$^{-1}$ (or equivalently, $\alpha \sim 10^{-37}$ GeV)
could be probed. Current upper bounds on this parameter from the
Super-Kamiokande experiment are approximately $\sim 10^{-23}$ GeV
\cite{lorentz,superk}.

\item{{\bf String Inspired Model}} \\
It has been suggested that $\alpha$ may scale with $E^2$,
particularly within the context of string theory
\cite{stringmotivation}. If this is the case, a 10
kiloparsecs distant source of TeV neutrinos could be used to test
$\kappa \sim 10^{-21}$ m$^{-1}$ TeV$^{-2}$, where $\kappa \equiv
\alpha/E^2$. At the energies observed by Super-Kamionkande ($\sim$GeV), this corresponds to $\alpha \sim 10^{-43}$ GeV. In this model, Super-Kamiokande's upper
limits are $\sim 10^{-10}$ GeV
\cite{lorentz,superk}.

\item{{\bf Lorentz Invariant Model}} \\
It has been shown that Lorentz invariance can be maintained if
$\alpha$ is proportional to $1/E$ \cite{lorentz}. In this case,
a 10 kiloparsecs distant source of TeV neutrinos could be
used to test $\mu^{2} \sim 10^{-21}$ m$^{-1}$ TeV, where $\mu^{2}
\equiv \alpha E$. At GeV energies, this corresponds to $\alpha
\sim 10^{-34}$ GeV. Super-Kamionkande's limit for this model is $\sim 10^{-22}$ GeV \cite{lorentz,superk}.

\end{itemize}

In each of these cases, neutrino studies over $\sim\,$10
kiloparsec baselines allow for tests of quantum decoherence at a
level many orders of magnitude beyond current bounds.

To measure the flavors of high-energy cosmic neutrinos, we turn to
large volume neutrino telescopes \cite{review}. In particular, the IceCube
experiment at the South Pole is currently under construction
\cite{icecube}. IceCube will be capable of observing both muon
tracks generated by charged current muon neutrino interactions and
hadronic or electromagnetic showers generated by charged current
electron neutrinos or neutral current interactions by all neutrino
flavors. With a full cubic kilometer of instrumented volume,
IceCube will be sensitive to neutrinos of energy as low as 100 GeV
and as high as $10^{11}$ GeV.  If constructed, a
kilometer-scale neutrino telescope in the Mediterranean could have
similar capabilities. 

The ability to measure the ratios of cosmic neutrinos in
high-energy neutrino telescopes has been studied in
Ref.~\cite{measure}. Such measurements have been discussed as
a test of neutrino stability \cite{decay}, pseudo-Dirac neutrinos
\cite{pseudodirac} and as a method to measure the neutrino mixing
angle, $\theta_{13}$ \cite{theta13}.

Assuming, for example, the neutrino flux calculated from
Cygnus in Ref.~\cite{halzenneutron} of $\sim 2 \times 10^{-11} \,
\rm{cm}^{-2} \rm{s}^{-1}$ integrated above 1 TeV, we can estimate
the ability of IceCube to distinguish the
$\nu_e$:$\nu_{\mu}$:$\nu_{\tau}=$0.33:0.33:0.33 and 0.56:0.24:0.20
cases. Practically, experiments do not measure the flux of each
neutrino flavor separately, but rather they measure the number of
(or spectrum of) muon events and shower events. This can, in turn,
be used to infer the fraction of neutrinos which are of muon type.
Based on the analysis of Ref.~\cite{measure}, it appears likely
that even after one year, this flux would produce enough events in
a kilometer-scale neutrino telescope to differentiate these two
cases. The precision of this technique would be further improved
by accumulating data over a period of several years.

\section{Conclusions}

In this letter, we have studied the possibility of using
high-energy neutrino telescopes to test for the effects of quantum
decoherence in cosmic neutrinos. Neutrinos produced in the decay
of charged pions produce ratios of neutrino flavors which are
indistinguishable from the signatures of decoherence after the
effects of oscillations are taken into account. Instead, we
consider high-energy cosmic neutrinos produced in the decay of
neutrons. With evidence accumulating for the presence of a neutron
component in the ultra-high energy cosmic ray flux, it is likely
that sizable electron anti-neutrino fluxes will also be present.
After oscillations, these neutrinos will reach Earth in the ratio
of $\nu_e$:$\nu_{\mu}$:$\nu_{\tau}=$0.56:0.24:0.20, in contrast to
the $\nu_e$:$\nu_{\mu}$:$\nu_{\tau}=$0.33:0.33:0.33 prediction for
a decohered flux. Next generation high-energy neutrino telescopes,
such as IceCube, should be capable of distinguishing these cases.
Neutrino flavor measurements over multi-kiloparsec baselines could
be used to place limits on the scale of quantum decoherence
between approximatley $10^{-34}$ and $10^{-43}$ GeV, depending on the model.
Considering the current bounds from the Super-Kamiokande
experiment are in the range of $\sim 10^{-10}$ to $10^{-23}$ GeV, it is clear that this
technique represents a major improvement in sensitivity.

\vspace{0.5cm}

{\it Acknowledgments}: We would like to thank Pedro Ferreira and Bob McElrath for helpful comments. DH is supported by the Leverhulme Trust. DM
is supported by the University of Sheffield. 

\end{document}